\documentclass[useAMS,usenatbib]{mn2e}
\usepackage{epsfig}
\title[Finite Larmor Radius Effects on  Dilute Plasmas]{Finite Larmor Radius Effects on Weakly Magnetized, Dilute Plasmas}
\author[Ebru Devlen and E. Rennan Pek\"{u}nl\"{u}]{Ebru Devlen$^{1}$\thanks{E-mail:
ebru.devlen@ege.edu.tr} and E. Rennan Pek\"{u}nl\"{u}$^{1}$\thanks{E-mail:
rennan.pekunlu@ege.edu.tr}\\
$^{1}$University of Ege, Faculty of Science, Department of Astronomy \& Space Sciences, Bornova, 35100, IZMIR, TURKEY}

\begin{document}

\date{Accepted . Received ; in original form }

\pagerange{\pageref{firstpage}--\pageref{lastpage}} \pubyear{}

\maketitle

\label{firstpage}

\begin{abstract}
We investigate the stability properties of a hot, dilute and differentially rotating weakly magnetized plasma which is believed to be found in the interstellar medium of galaxies and protogalaxies and in the low-density accretion flows around some giant black holes like the one in the Galactic center. In the linear MHD regime, we consider the combined effects of gyroviscosity and parallel viscosity  on the magnetorotational instability. The helical magnetic field is considered in the investigation. We show that the gyroviscous effect and the pitch angles cause a powerful gyroviscous instability. Furthermore, in most of the cases, plasma with the above mentioned properties is unstable and the growth rates of the unstable modes are larger than that  of  the magnetorotational instability. 
\end{abstract}


\section{Introduction}

Enhanced outward angular momentum transport problem in accretion discs have been treated in hydrodynamic (HD) and magnetohydrodynamic (MHD) contexts by referring to the fluid description as well as the kinetic approximation. An overview of the past efforts on solving the problem is given by Balbus \& Hawley (1998) and Balbus (2003). 

Shakura and Sunyaev (1973) appealed to the large Reynolds number to get over the difficulties presented by inadequate kinematic viscosity. The breakthrough eventually was made by Balbus and Hawley (1991) when they showed that differentially rotating an accretion disc with angular velocity decreasing outward and threaded by a weak magnetic field is linearly unstable. This instability is known as the magnetorotational instability (MRI). Balbus and Hawley (1998) showed that even non-Keplerian ``thick"  discs are unstable in the presence of a weak magnetic field  and thus claimed the more generality of the MRI. The literature concerning the applicability of the MRI in linear as well as nonlinear regimes on various astrophysical systems is rich in content. The reader may refer to the review article by Balbus (2003).

Balbus (2004) investigated the instability of  the differentially rotating dilute plasma 
and showed that the maximum growth rate of the magnetoviscous instability is greater than that of the MRI. Also, he noted that the parallel viscous stress can lead to accretion disk turbulence, even if Lorentz force is negligible.
Balbus refers to the resulting viscous tensor, in the limit of $\omega_{ci}\tau_{i}\gg 1$ , as the Braginskii viscosity which he claims to be appropriate for interstellar, galactic and protogalactic discs. But Balbus (2004) did not take gyroviscous force which is one of the components of the stress tensor and due to finite Larmor radius (FLR) effect  into account. 

Quataert, Dorland and Hammett (2002) explored the kinetic version of the MRI which is the regime where the wavelengths are much larger than the proton Larmor radius in a hot accretion flows onto compact objects. They showed that in the various limits of plasma beta, growth rates differ considerably. The major conclusion they draw is that kinetic effects they considered do not change the stability criterion from the MHD result, but do change the growth rate significantly at high beta.

Ramos (2003) presented an analysis on the dynamic evolution of the parallel heat fluxes in a collisionless magnetized plasma. Fluid description of collisionless plasmas was first given by Chew, Goldberger and Low (1956). They worked in the lowest order or zero Larmor radius limit. Diamagnetism and other multifluid effects require higher orders in the gyroradius expansion. Taking into account the above mentioned shortcomings facing the collisionless magnetised plasmas, Ramos (2005) derived the first-significant-order FLR systems of fluid moment equations. He claims that this formalism can account for the gyroviscous stress, the pressure anisotropy and the anisotropic heat fluxes in a plasma with an arbitrary magnetic field geometry.

Islam and Balbus (2005) generalized the viscous instability examined in Balbus's earlier study (Balbus 2004) by including the dynamical effects of magnetic tension force. They found that the growth rates are lower than the case when tension force is not considered. Nevertheless, they argue, the growth rates are still higher than the maximum of the standart MRI. They conclude that the magnetoviscous instability may find an application in galactic discs and halos and low density accretion flows around the Galactic center.

Ferraro (2007) considered the hitherto neglected ion gyroviscosity effect which represents the first-order FLR  corrections to the two-fluid MHD equations. He showed that FLR effects are much more important than the Hall effect  in  the limit of weak magnetic fields. Also, Ferraro claimed that in the collisional limit, gyroviscous effects at scales much larger than $r_{Li}$ may completely stabilizes the MRI. But he considered the magnetic field  configuration with $\mathbf{B}_{0}=B\mathbf{\hat z}$ that the parallel viscosity doesn't  play a role in the MRI. Although it is clear that  extending his analysis to a more general magnetic field configuration (especially that the parallel viscosity is important)  complicates the analysis considerably, it is necessary to understand the true nature of this instability. Also, FLR effects that produces collisionless viscosity represent ``non-ideal" MHD terms  like Hall term in the fluid formalism. Therefore, the FLR effects  may  introduce important corrections to the dispersion relations for MHD instabilities and waves. An investigation considering both parallel viscosity and gyroviscous effects is the primary motivation of  this paper.

The plan of the paper is as follows: in Section 2, we give our two fluid equations containing hithereto unexplored hot, dilute differentially rotating plasma medium threaded by a weak (subthermal, i.e. magnetic energy density is small compared with the thermal energy density) helical magnetic field wherein the gyroviscosity and parallel viscosity may play important roles in stability. In Section 3 we present the linearized MHD equations and their solutions and finally in Section 4 discussion and conclusion are given.

\section{TWO FLUID EQUATIONS}

In a dilute plasma, ion gyroradius  is much smaller than the ion collision mean free path, i.e., $r_{Li}\ll\lambda_{i}$,  and also ion cyclotron frequency ($\omega_{ci}$) greatly exceeds ion-ion collision frequency ($\nu_{i}$), i.e., $\epsilon\equiv\omega_{ci}\tau_{i}\gg 1$ where $\tau_{i}=1/\nu_{i}$. Conventional MHD theory assumes that  plasma pressure is isotropic for simplicity. However, the one that want to make a more accurate model of plasma must take into consideration anisotropy in the pressure. The anisotropy in the pressure can be due to the presence of a strong magnetic field under laboratory conditions, but under the astrophysical conditions, the plasma may have a tensorial character even in  the presence of a  weak magnetic field, the resulting anisotropy being determined by the Larmor frequencies and the macroscopic velocity gradients of the  constituents of the plasma. FLR effects are accounted for by the gyroviscosity, which are terms in the pressure tensor independent of collision frequency(Roberts \& Taylor 1962). 

Spitzer (1962) gives $\epsilon$ parameter for a hydrogenic plasma as below:

\begin{equation}
\epsilon=\left({{1.09\times10^5}\over{n}}\right) {T_4^{3/2}B_{\mu G}\over\ln\Lambda}
\end{equation} 
where $n$ is the proton density in cm$^{-3}$, $T_4$ the temperature in units of $10^4$ K, $B_{\mu G}$ is the magnetic field in microgauss and $\ln\Lambda$ is the Coulomb logarithm. For $n\la1$ and $T_4\ga1$, the condition of $\epsilon\gg1$ is fulfilled even in the presence of a  very weak field. Also Islam \& Balbus (2005) mentioned the one which assumes large Reynolds number will obtain that  the collision frequency is much larger than an orbital frequency but much smaller than the cyclotron frequency (i.e., $\Omega\ll\nu_{i}\ll\omega_{ci}$).

Under this condition, MHD equations describing the plasma dynamics should include the anisotropic transport terms arising from the free flow of the particles along the magnetic field lines (Braginskii 1965). The parallel viscosity  by ions is higher than the electrons  by the factor $(m_{i}/m_{e})^{1/2}$. In order to take into account the arbitrarily high order FLR effects the MHD equations should contain collisionless viscous tensor. Balbus (2004) stressed on the importance of ion viscosity in a rotating systems. In an attemp to investigate the FLR effect and parallel viscosity in a dilute plasma one should consider two fluid equations. The equilibrium state is a differentially rotating dilute plasma. 

The two-fluid moment equations together with Faraday and Ampere laws, respectively are given as below (Braginskii 1965):

\begin{equation}
{dn_{s}\over dt}=-n_{s}\nabla\cdot\mathbf{v}_{s}
\end{equation}

\begin{equation}
m_{s}n_{s}{d{\mathbf{v}_{s}}\over dt}=-\nabla P_{s}-\nabla\cdot\mathbf{\Pi}_{s}+q_{s}n_{s}\left(\mathbf{E}+{\mathbf{v}\times\mathbf{B}\over c}\right)-m_{s}n_{s}g
\end{equation}

\begin{equation}
{\partial{\mathbf{B}}\over\partial{t}}=-c\nabla\times\mathbf{E}
\end{equation}

\begin{equation}
{\mathbf{J}}={c\over 4\pi}\nabla\times{\mathbf{B}}=en_{e}(\mathbf{v}_{i}-\mathbf{v}_{e})
\end{equation}

where the subscript ``s" stands for electrons and ions. The symbols have their usual meanings. $d/dt=\partial/\partial t+(\mathbf{v}\cdot\nabla)$ is the Lagrangian derivative. 

If we assume quasi-neutrality, i.e., $n_{e}=Zn_{i}$ then the mass continuity equation is written only for ions. Let us put $P=P_{i}+P_{e}, \mathbf{\Pi}=\mathbf{\Pi}_{i}+\mathbf{\Pi}_{e}$ and assume that $m_{e}/m_{i}\sim 0$. With these substitutions the MHD equations become,

\begin{equation}
{d\rho\over dt}+\rho\nabla\cdot\mathbf{v}=0
\end{equation}

\begin{equation}
\rho{d{\mathbf{v}}\over dt}=-\nabla P-\nabla\cdot\mathbf{\Pi}+{\mathbf{J}\times\mathbf{B}\over c}-\rho g(R)
\end{equation}

\begin{equation}
{\partial{\mathbf{B}}\over\partial{t}}=\nabla\times \left[\mathbf{v}\times\mathbf{B}-{\mathbf{J}\times\mathbf{B}\over en_{e}}+{c\nabla P_{e}\over en_{e}}+{c\nabla \cdot\mathbf{\Pi}_{e}\over en_{e}}\right]
\end{equation}

Stress tensor $\mathbf{\Pi}=\mathbf{\Pi}^{\parallel}+\mathbf{\Pi}^{\perp}+\mathbf{\Pi}^{gv}$ is described as the sum total of the parallel $(\parallel)$, perpendicular $(\perp)$  and the gyroviscous $(gv)$  components (Braginskii, 1965). In a dilute plasma perpendicular viscosity is smaller than the parallel viscosity by a factor $\left(r_{L}/\lambda\right)^{2}$ where $r_{L}$ is the Larmor radius and $\lambda$ is the mean free path of the particles. $\mathbf{\Pi}^{gv}$ is a dissipationless stress which represents the lowest-order FLR correction to the fluid equations and measures the changes of particle drift velocities across a gyroorbit (Ferraro 2007). We use the Braginskii closure for the stress tensor $\mathbf{\Pi}=\mathbf{\Pi}^{v}+\mathbf{\Pi}^{gv}$ in our investigation (see, e.g., Ferraro 2007).

\begin{equation}
\mathbf{\Pi}^{v}=0.96{P_{i}\over 2\nu_{i}}\left(\mathbf{I}-3\mathbf{\hat{b}}\mathbf{\hat{b}}\right)\left(\mathbf{\hat{b}}\cdot\mathbf{W}\cdot\mathbf{\hat{b}}\right)
\end{equation}

\begin{equation}
\mathbf{\Pi}^{gv}={P_{i}\over 4\omega_{ci}}\left[\mathbf{\hat{b}}\times\mathbf{W}\cdot\left(\mathbf{I}+3\mathbf{\hat{b}}\mathbf{\hat{b}}\right)+\left[\mathbf{\hat{b}}\times\mathbf{W}\cdot\left(\mathbf{I}+3\mathbf{\hat{b}}\mathbf{\hat{b}}\right)\right]^{T}\right]
\end{equation}

Here   $ \mathbf{\hat{b}}=\mathbf{B}/B$, ${\omega_{ci}}=eB/m_{i}c$  are the unit vector along the magnetic field and the cyclotron frequency, respectively; $ \nu_{i}$ ion collision frequency. $\mathbf{W}=\nabla\mathbf{v}+(\nabla\mathbf{v})^{T}-2/3\mathbf{I}\left(\nabla\cdot\mathbf{v}\right)$ is the rate of strain tensor.

\section{LINEARIZED EQUATIONS}

We assume that the Hall effect which is represented by the second term on the right hand side of the equation (8) is negligible for $\beta\gg1$ (Ferraro 2007). This effect is the most important one in the low-$\beta$ ($\beta$ is the ratio of the gas pressure to magnetic pressure.) plasmas.  Therefore will not be considered in the present investigation. The third term on the right hand side of the equation (8) is the thermodiffusion term. Due to their higher masses, ions carry the most of the momentum. Therefore the last term on the right hand side of the equation (8) is negligible.  We work in the Boussinesq limit and set $\nabla\cdot\mathbf{v}=0$. 
The velocity $\mathbf{v}$ in the equations (6)--(8) is the ion velocity. We seek the solution of the dispersion relation to be derived from the set of equations (6)--(8). We work in a cylindrical coordinate system, $(R, \phi, z)$. Space-time dependence of the axisymmetric perturbations is assumed to be of the form, $\exp(ik_{R}+ik_{z}+\omega{t})$. The weak magnetic field has a helical shape with $B_{\phi}=B_{0}\cos{\theta}$ and $B_{z}=B_{0}\sin{\theta}$, where $\theta=\tan^{-1}(B_{z}/B_{\phi})$ is the angle between the magnetic field vector and the $\phi$ axis of the coordinate system. The \textit{pitch angle} we define as the one between the $\phi$ direction of the cyclindrical coordinate system and the magnetic field \textbf{B}, not as the angle between the instantaneous velocity vector and the magnetic field as defined in the \textit{particle orbit theory}.   $B_{0}$ is the magnitude of the seed field. The reason why we assumed  $B_{R}=0$ is that the presence of a finite $B_{R}$ generates a time dependent $B_{\phi}$ (Balbus \& Hawley 1991) which in turn makes the analysis complicated. In the equilibrium state of the differentially rotating plasma with a Keplerian velocity profile, $ v_{\phi}=R\Omega(R)$, plasma pressure is assumed to be isotropic.

The linearized equation of mass continuity is as below:

\begin{equation}
k_{R}\delta v_{R}+k_{z}\delta v_{z}=0
\end{equation}

The radial, azimuthal and the axial components of the linearized momentum conservation equation are given by the equations (12), (13) and (14), respectively, 
\begin{eqnarray}
\omega\delta{v}_{R}-2\Omega\delta{v}_{\phi}+ik_{r}{\delta P\over\rho}\nonumber\\
+{1\over4\pi\rho}ik_{R}\left({B}_{\phi}\delta{B}_{\phi}+
{B}_{z}\delta{B}_{z}\right)
-{1\over4\pi\rho}ik_{z}{B}_{z}\delta{B}_{R}\nonumber\\
-V_{par}\left[\begin{array}{c}{d\Omega\over d\ln R}{1\over\omega}{k_{R}\over k_z}\sin2\theta\delta v_{R}+{k_{R}\over k_z}\sin2\theta \delta v_{\phi}\\
+2{k_{R}\over k_z}\sin^{2}\theta\delta v_{z}
\end{array}
\right]\nonumber\\
-V_{gyro}\left[\begin{array}{c}
2\sin\theta{d\Omega\over d\ln R}{1\over\omega}({k_{R}^2\over k_{z}^2}-1)\delta v_{R}\\
+(2\cos\theta {k_{R}^2\over k_{z}^2}+E)\delta v_{z}\\
-(2{k_{R}^2\over k_{z}^2}\sin\theta+A)\delta v_{\phi}
\end{array}
\right]=0
\end{eqnarray}

\begin{eqnarray}
\omega\delta{v}_{\phi}+{\kappa^{2}\over 2\Omega}\delta{v}_{R}-{1\over4\pi\rho}ik_{z}\left({B}_{z}\delta{B}_{\phi}\right)\nonumber\\
+V_{par}2D\left[{d\Omega\over d\ln R}{1\over\omega}cos\theta\delta v_{R}+cos\theta \delta v_{\phi}+\sin\theta\delta v_{z}\right]\nonumber\\
+V_{gyro}\left[\begin{array}{c}
-2D{d\Omega\over d\ln R}{1\over\omega}
\delta v_{z}
+{d\Omega\over d\ln R}{1\over\omega}B\delta v_{\phi}\\
+\left[({d\Omega\over d\ln R})^2{1\over\omega^{2}}B
-A+2\sin\theta {k_{R}^2\over k_{z}^2}\right]\delta v_{R}
\end{array}
\right]=0
\end{eqnarray}

\begin{eqnarray}
\omega\delta{v}_{z}+ik_{z}{\delta P\over\rho}+{1\over4\pi\rho}ik_{z}{B}_{\phi}\delta{B}_{\phi}\nonumber\\
+V_{par}\left[F {d\Omega\over d\ln R}{1\over\omega}
\delta v_{R}+F\delta v_{\phi}+C\delta v_{z}\right]\nonumber\\
+V_{gyro}\left[\begin{array}{c}
\left(2{k_{R}^2\over k_{z}^2}\cos\theta+E+4D({d\Omega\over d\ln R})^2{1\over\omega^{2}}\right)\delta v_{R}\\
+\left(4{k_{R}\over k_{z}}\sin\theta+4D{d\Omega\over d\ln R}{1\over\omega}\right)\delta v_{\phi}\\
+{d\Omega\over d\ln R}{1\over\omega}G\delta v_{z}
\end{array}
\right]=0
\end{eqnarray}

and similarly, the radial, azimuthal and axial components of the linearized magnetic induction equation are given by the equations (15), (16) and (17) respectively,

\begin{equation}
\omega\delta B_{R}-ik_{z}B_{z}\delta v_{R}=0
\end{equation}
\begin{equation}
\omega\delta B_{\phi}-ik_{z}B_{z}\delta v_{\phi}-{d\Omega\over d\ln R}\delta B_{R}=0
\end{equation}
\begin{equation}
\omega\delta B_{z}-ik_{z}B_{z}\delta v_{z}=0
\end{equation}

where $V_{par}=0.96k_{z}^2P_{i}/2\nu_{i}\rho$ is the inverse time scale of the dissipation due to parallel viscosity; $V_{gyro}=k_{z}^2P_{i}/4\omega_{ci}\rho$ is the inverse time scale of the gyroviscous stress.  $\kappa^2$ is epicyclic frequency. Other constants which depends on pitch angle $\theta$ are $A=\sin\theta(1-3\cos2\theta)$, $B=\sin\theta(1+9\cos^2\theta-3\sin^2\theta)$, $C=2\sin^2\theta(-1+3\sin^2\theta)$, $D= 3\sin^2\theta\cos\theta$, $E=2\cos\theta(1+3\sin^2\theta)$, $F=2\sin\theta D-\sin2\theta$, $G=\sin\theta(1+3\cos 2\theta)$ and $H=3\sin\theta\cos^2\theta$.

The  dispersion relation (DR) derived from the set of equations (11)-(17) is given below: 

\begin{equation}
{\omega}^{4}+{a}_{3}{\omega}^{3}+{a}_{2}{\omega}^{2}+{a}_{1}{\omega}+{a}_{0}=0
\end{equation}

where \begin{equation}
{a}_3=6{V}_{par}s^2{k_{\perp}^2\over k^2}
\end{equation}

\begin{eqnarray}
{a}_2=2k_{z}^2v_{A}^{2}+{k_{z}^2\over k^2}{\kappa}^2+{k_{z}^2\over k^2}\tilde{V}_{gyro}^{2}{\Omega}^2(2sy^2-A)^2\nonumber\\
+\tilde{V}_{gyro}\Omega^{\prime}(G+H)+4{k_{z}^2\over k^2}\tilde{V}_{gyro}(2sy^2-A)
\end{eqnarray}

\begin{eqnarray}
{a}_1=6{V}_{par}s^2k_{z}^2v_{A}^{2}{k_{\perp}^2\over k^2}+{V}_{par}\tilde{V}_{gyro}3s^2G\Omega^{\prime}{k_{\perp}^2\over k^2}\nonumber\\
+{k_{z}^2\over k^2}{V}_{par}\Omega^{\prime}2cD+{k_{z}^2\over k^2}\tilde{V}_{gyro}6y\Omega D\Omega^{\prime}
\nonumber\\
+(\tilde{V}_{gyro})^2 3y\Omega D(2sy^2-A)\Omega^{\prime}{k_{z}^2\over k^2}
\end{eqnarray}

\begin{eqnarray}
{a}_0={k_{z}^2\over k^2}y^2\tilde{V}_{gyro}^2 2D^2(\Omega^{\prime})^2+\left[k_{z}^2v_{A}^{2}+\tilde{V}_{gyro}\Omega^{\prime}(G/2+H)\right]\nonumber\\
\times\left[k_{z}^2v_{A}^{2}+\tilde{V}_{gyro}\Omega^{\prime}G/2+{k_{z}^2\over k^2}\Omega^{\prime}\right]
\end{eqnarray}

where $y=k_R/k_z$, $\Omega^{\prime}=d\Omega^2/d\ln R$, $s=sin\theta$, $c=cos\theta$, $k_{\perp}^2=k_{R}^2+k_{z}^2\cos^2\theta $ and $\tilde{V}_{gyro}=k_{z}^2P_{i}/4\Omega\omega_{ci}\rho$.
If we set  $\mathbf{\Pi}$ the stress tensor to zero and assume $\theta=0$, our dispersion relation is reduced to the one given by Balbus \& Hawley (1991) (taking into account the difference in their definition of omega; see their equation 2.9). If we set the gyroviscous force to zero, then we recover the dispersion relation given by Islam \& Balbus (2005) (see their equation 31).	

\begin{figure*}
\includegraphics[scale=1.1]{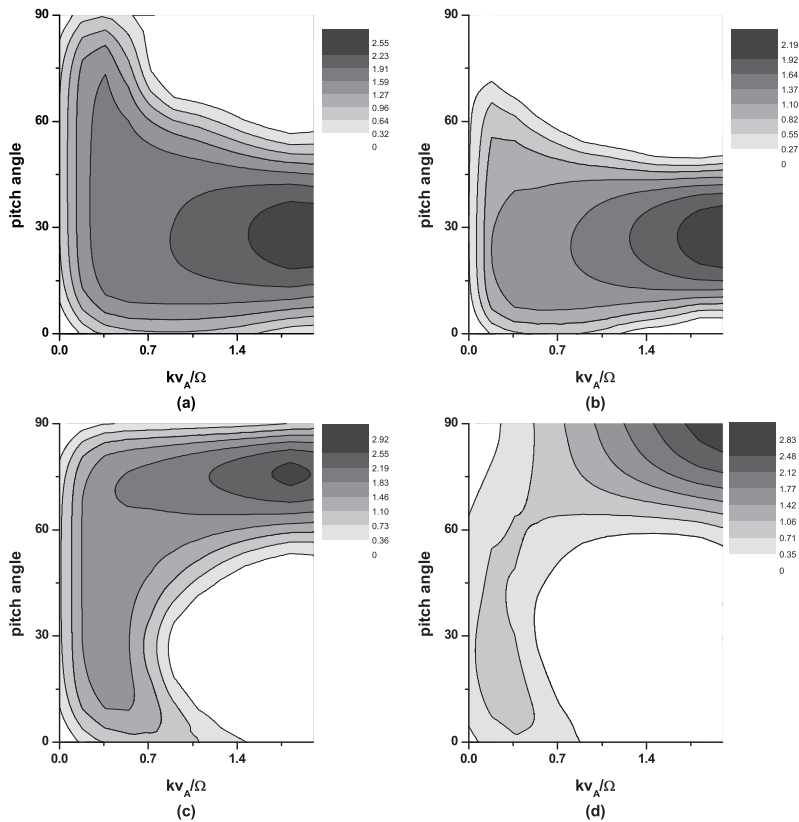}
%\includegraphics[]{fig1.eps}
%\epsscale{.60}
%\vspace*{174pt}
\caption{The growth rate of gyroviscous instability for the cases of $\mathbf{\Omega}\uparrow\uparrow\mathbf{B}_{z}$ (Fig. 1a and 1b) and $\mathbf{\Omega}\uparrow\downarrow\mathbf{B}_{z}$ (Fig. 1c and 1d). Figure 1a and 1c is drawn for $k_{R}=0$ and Figure 1b and 1d for $k_{R}/k_{z}=1$. In the situation  $\tilde{V}_{gyro}^{n}=1$ for the pitch angles are smaller than $60^{\circ}$ and in the situation  $\tilde{V}_{gyro}^{n}=-1$ for the pitch angles are greater than $60^{\circ}$ the instability spreads over  a wider range of wavenumbers.} 
\label{fig1}
\end{figure*}

\subsection{Instability Criterion and Numerical Solutions}

Equation (18) describes four low-frequency modes that exist in a plasma properties of which described above.  A necessary and sufficient criterion for instability is clearly given by the Routh-Hurwitz theorem

\begin{eqnarray}
a_{0}={k_{z}^2\over k^2}y^2\tilde{V}_{gyro}^2 2D^2(\Omega^{\prime})^2+\left[k_{z}^2v_{A}^{2}+\tilde{V}_{gyro}\Omega^{\prime}(G/2+H)\right]\nonumber\\
\times\left[k_{z}^2v_{A}^{2}+\tilde{V}_{gyro}\Omega^{\prime}G/2+{k_{z}^2\over k^2}\Omega^{\prime}\right]<0
\end{eqnarray}

In the absence of gyroviscosity effect,  this criterion is recovered as $(\mathbf{k}\cdot\mathbf{v_A})^2<-\Omega^{\prime}$ and it is an ideal MRI  criterion 
which states ``the combination of a negative angular velocity radial gradient with almost any small field will lead to dynamical instability" (Balbus \& Hawley 1991). 
When we consider FLR effect the instability criterion (23) is very complicated. Because it depends on magnetic tension force, gyroviscosity force, angular velocity gradient  and $\theta$-pitch angle. 
Also, angular velocity gradient, gyroviscosity force and constants depend on pitch angle may be positive or negative. For the simplicity we consider only $\mathbf{B}=B_{0}\mathbf{\hat z}$ case. In this situation pitch angle is $90^\circ$ and $D=0$, $G=-2$, $H=0$. If we assume $\Omega^{\prime}<0$, the instability  depends on the sign of $\tilde{V}_{gyro}$. For the case of $\tilde{V}_{gyro}>0$, since the first factor of the second term is positive, the instability criterion is reduced to the  condition that the  second factor must be  negative. Although gyroviscous force is coupled with differential rotation which is the source of free energy of MRI, it acts in the same direction as the magnetic tension force. Therefore it suppresses the instability. For the case of $\tilde{V}_{gyro}<0$, the gyroviscous force acts in the opposite direction of the magnetic tension force and enhances the instability. On the other hand, for $\Omega^{\prime}>0$,  the situation is reversed. As a result, the gyroviscosity force stabilizes or destabilizes according to whether $(\mathbf{\Omega}\cdot\mathbf{B})$ is positive or negative. In this regard, the � FLR effect is similar to the Hall effect. The contribution of the first term (in the inequality (23)) to the instability arises only when $k_{R}$ is taken into account and $0^\circ<\theta<90^\circ$. This term is always positive, therefore it has a stabilizing effect. When we consider different pitch angles and wavenumbers, the instability criterion is highly complicated due to the change of the signs and relative magnitudes of the terms. Therefore, in order to reach a definitive results the numerical solutions of the dispersion relation (18) are required. We have rewrittten the dispersion relation in its  dimensionless form, i.e. all the terms of Eqn.18 is divided by $\Omega^4$:

\begin{equation}
{\gamma}^{4}+{b}_{3}{\gamma}^{3}+{b}_{2}{\gamma}^{2}+{b}_{1}{\gamma}+{b}_{0}=0
\end{equation}

where \begin{equation}
{b}_3=6\tilde{V}_{par}^{n}Xs^2{k_{\perp}^2\over k^2}
\end{equation}

\begin{eqnarray}
{b}_2=\tilde{V}_{gyro}^{n}X{d\ln\Omega^2\over d\ln R}(G+H)+4{k_{z}^2\over k^2}\tilde{V}_{gyro}^{n}X(2sy^2-A)\nonumber\\
+2X+{k_{z}^2\over k^2}\tilde{\kappa}^2+{k_{z}^2\over k^2}(\tilde{V}_{gyro}^{n})^{2}X^2(2sy^2-A)^2
\end{eqnarray}

\begin{eqnarray}
{b}_1=6\tilde{V}_{par}^{n}s^2k_{z}^2X^2{k_{\perp}^2\over k^2}+\tilde{V}_{par}^{n}\tilde{V}_{gyro}^{n}X^2 3s^2G{d\ln\Omega^2\over d\ln R}{k_{\perp}^2\over k^2}\nonumber\\
+{k_{z}^2\over k^2}\tilde{V}_{par}^{n}X{d\ln\Omega^2\over d\ln R}2cD+{k_{z}^2\over k^2}\tilde{V}_{gyro}^{n}6yX D{d\ln\Omega^2\over d\ln R}
\nonumber\\
+(\tilde{V}_{gyro}^{n})^2X^2 3y D(2sy^2-A){d\ln\Omega^2\over d\ln R}{k_{z}^2\over k^2}
\end{eqnarray}

\begin{eqnarray}
{b}_0=\left[X\left(1+\tilde{V}_{gyro}^{n}{d\ln\Omega^2\over d\ln R}G/2\right)+{k_{z}^2\over k^2}{d\ln\Omega^2\over d\ln R}\right]\nonumber\\
\times\left[X\left(1+\tilde{V}_{gyro}^{n}{d\ln\Omega^2\over d\ln R}(G/2+H)\right)\right]\nonumber\\
+{k_{z}^2\over k^2}y^2(\tilde{V}_{gyro}^{n})^2X^2 2D^2({d\ln\Omega^2\over d\ln R})^2
\end{eqnarray}

where $\gamma=\omega/\Omega$, $X=k_{z}^2v_{A}^{2}/\Omega^2$, $\tilde{V}_{gyro}^{n}=\tilde{V}_{gyro}/X$ and $\tilde{V}_{par}^{n}={V}_{par}/\Omega X$.

$\tilde{V}_{par}^{n}\gg \tilde{V}_{gyro}^{n}$ relationship is easily derived 
from condition of dilute plasma $\epsilon\equiv\omega_{ci}\tau_{i}\gg 1$ where $\tau_{i}=1/\nu_{i}$. Taking into consideration this relationship, all the figures are drawn as dimensionless growth rate versus wavenumber for  $\tilde{V}_{par}^{n}=1000$ and for the Keplerian rotational profile. The growth rate depends sensitively on the pitch angle and the gyroviscosity parameter $V_{gyro}/\Omega X=\tilde{V}_{gyro}^{n}$.

Figure 1a shows the fastest growing mode (i.e., $k_{R}=0$) in the case when the rotation axis ($\mathbf{\Omega}$) and the $\mathbf{B}_{z}$ are parallel ($\mathbf{\Omega}\uparrow\uparrow\mathbf{B}_{z}$); and Figure 1b is plotted for $k_{R}/k_{z}=1$ and also $\mathbf{\Omega}\uparrow\uparrow\mathbf{B}_{z}$. Figure 1a and 1b are drawn for  the $\tilde{V}_{gyro}^{n}=1$.  Although all the wavenumbers are unstable for the pitch angles  $15^\circ$ and $45^\circ$ when $k_{R}=0$ and $k_{R}/k_z=1$, for the $\theta>60^\circ$  all the wavenumbers are stable only when $k_{R}/k_z=1$.
Also, the growth rates  exceed the maximum growth rate of MRI ($0.75\Omega$) especially for the $\theta=30^\circ$ in the both cases.
When the pitch angle between the magnetic field lines and the angular velocity gradient is  $90^\circ$ there is the minimum viscous transport of angular momentum (Fig. 1a). Instability tends to grow when the angular velocity gradient acquires a component along the magnetic field lines. But for greater pitch angles ($>45^\circ$) the radius of curvature becomes greater, therefore tension force becomes smaller. In a dilute plasma, the dynamic effects caused by the magnetic tension force would lower the growth rates (Islam \& Balbus 2005). One would expect greater growth rates when the tension force is smaller. But, as Islam \& Balbus(2005) states that the dependence of the growth rate on the tension force (in our opinion, possibly through the pitch angle) shows the complicated nature of the magnetic stresses ``as both a stabilizing and a destabilizing agent". Therefore it seems as though the viscous angular momentum is more effectively transported when the pitch angle is within the range of about $15^\circ-45^\circ$.

%When we consider the parallel viscosity effect together with the gyroviscosity, the growth rates are lower than the case without parallel viscosity (Figures 1c and 1d).Figure 1c and 1d are drawn for the cases of   $k_{R}=0$ and $k_{R}/k_z=1$, respectively. Especially in Figure 1d the growth rate of the case of $60^\circ$ is clearly seen because wavevector $k_{R}$ is also taken into consideration.

When the angular velocity vector and the $B_{z}$ component of the magnetic field are anti parallel ($\mathbf{\Omega}\uparrow\downarrow\mathbf{B}_{z}$),  gyroviscosity assumes negative values. Figure 1c and 1d are drawn only for the  $\tilde{V}_{gyro}^{n}=-1$ in the  cases of   $k_{R}=0$ and $k_{R}/k_z=1$, respectively. In the case of $k_{R}=0$ all the wavenumbers are unstable only for the $\theta>45^\circ$ (Fig. 1c), in the case of $k_{R}/k_z=1$, they are unstable for the $\theta>60^\circ$ (Fig. 1d). In addition, in Figure 1c and 1d maxiumum growth rates are larger  than the ideal MRI growth rates for the great values of pitch angle. Besides, when we consider finite $k_{R}$ wavevector, an unstable mode emerges for $\theta=90^\circ$

%When we considered the parallel viscosity and the gyroviscosity, the results of the analysis did not change for  the case of   $k_{R}=0$, but the case of $k_{R}/k_{z}=1$ exhibits a different  character. The growth rates are decreased by the parallel viscosity for the $45^\circ$ and $60^\circ$, also the stable wavenumbers is destabilized by this effect for the $30^\circ$ and $90^\circ$.

The normalized wavenumber of the fastest growing mode is given as $k_{z}v_{A}/\Omega=(15/16)^{1/2}$ (Balbus \& Hawley 1992). We adopt the same value in the present investigation. We plotted dimensionless growth rates versus pitch angle and  $k_{R}/k_{z}$ in Figure 2.  For the $\tilde{V}_{gyro}^{n}=1$, growth rates are given in Figure 2a and for the  $\tilde{V}_{gyro}^{n}=-1$ they are given in Figure  2b.  In the case of  $\mathbf{\Omega}\uparrow\uparrow\mathbf{B}_{z}$ 
growth rates have maximum values for the  small pitch angles and $k_{R}$ values (Fig 2a). While there is no unstable wavenumber for the $\theta>60^\circ$ (Fig 2a) for the case of  $\mathbf{\Omega}\uparrow\downarrow\mathbf{B}_{z}$ these pitch angles have maximum growth rates (Fig 2b). 

To see the  effect of the gyroviscosity parameter on the growth rates  we assumed the  pitch angle as $\theta=45^\circ$. Then,  we plotted dimensionless  growth rates versus gyroviscosity parameter and wavenumbers for the $k_{R}/k_z=0$, $k_{R}/k_z=1$ and $k_{z}v_{A}/\Omega=(15/16)^{1/2}$ respectively in the Figure 3a, 3b and 3c. We clearly see that all wavenumbers are unstable for the positive values of the groviscosity parameter, i.e. in the case of $\mathbf{\Omega}\uparrow\uparrow\mathbf{B}_{z}$.  The positive values of the gyroviscosity parameter have maximum growth rates case in the large wavenumbers, but again the growth rates are greater than $0.75 \Omega$.

In Figure 4, we assumed the  pitch angle as $\theta=60^\circ$. Again,  we plotted dimensionless  growth rates versus gyroviscosity parameter and wavenumbers for the $k_{R}/k_z=0$, $k_{R}/k_z=1$ and $k_{z}v_{A}/\Omega=(15/16)^{1/2}$ respectively in the Figure 4a, 4b and 4c. We clearly see that all wavenumbers are unstable for the negative values of the groviscosity parameter, i.e. in the case of $\mathbf{\Omega}\uparrow\downarrow\mathbf{B}_{z}$.  Maximum growth rates emerge in the small wavenumbers and  and the negative values of the gyroviscosity parameter contrary to previous case (Figure 3).

\begin{figure*}
\includegraphics[scale=1.3]{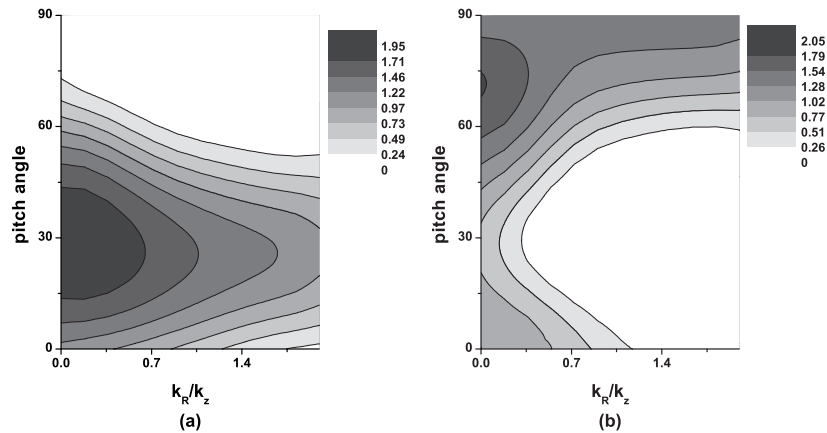}
\caption{The dimensionless growth rates of instability versus pitch angles and $k_{R}/k_{z}$ for the normalized wavenumber of the fastest growing mode, i.e. $k_{z}v_{A}/\Omega=(15/16)^{1/2}$. 
In the situation  $\tilde{V}_{gyro}^{n}=1$ (Fig. 2a) for the pitch angles are smaller than $60^{\circ}$ and in the situation  $\tilde{V}_{gyro}^{n}=-1$ (Fig. 2b) for the pitch angles are greater than $60^{\circ}$ all the wavenumbers are unstable.}
\label{fig2}
\end{figure*}

\begin{figure*}

\hspace*{-2cm}
\includegraphics[scale=2]{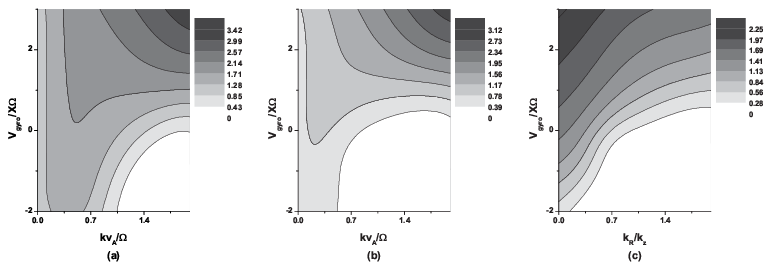}

%\epsscale{.80}
%\vspace*{174pt}
\caption{The dimensionless  growth rates of instability versus gyroviscosity parameter and wavenumbers for the $k_{R}/k_z=0$, $k_{R}/k_z=1$ and $k_{z}v_{A}/\Omega=(15/16)^{1/2}$ respectively in the Figure 3a, 3b and 3c. The pitch angle is taken $45^{\circ}$  in these figures. For the positive values of gyroviscosity parameter, i.e. in the case of $\mathbf{\Omega}\uparrow\uparrow\mathbf{B}_{z}$, all the wavenumbers are unstable.}
\label{fig3}

\end{figure*}

\begin{figure*}
\hspace*{-2cm}
\includegraphics[scale=2]{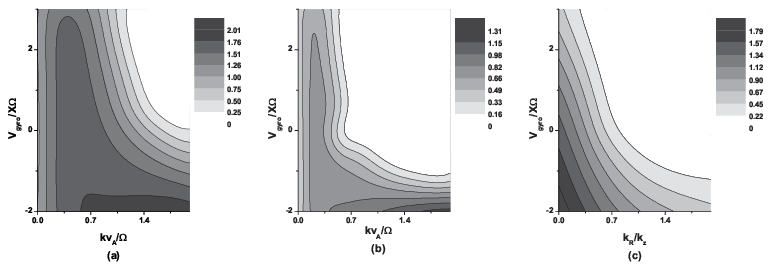}
%\epsscale{.80}
%\vspace*{174pt}
\caption{The dimensionless  growth rates of instability versus gyroviscosity parameter and wavenumbers for the $k_{R}/k_z=0$, $k_{R}/k_z=1$ and $k_{z}v_{A}/\Omega=(15/16)^{1/2}$ respectively in the Figure 4a, 4b and 4c. The pitch angle is taken $60^{\circ}$  in these figures. For the negative values of gyroviscosity parameter, i.e. in the case of $\mathbf{\Omega}\uparrow\downarrow\mathbf{B}_{z}$, all the wavenumbers are unstable.}
\label{fig4}
\end{figure*}

\section{Conclusions}
In this paper we  have investigated a linear axisymmetric analysis of the MRI with FLR effect and paralel viscosity in the hot, dilute and differentially rotating weakly magnetized plasma.  In a dilute plasma, ion cyclotron frequency  much exceeds the  ion-ion collision frequency. In this regime, the viscous stress tensor is given as the total of the parallel and the gyroviscous components (Braginskii 1965). This work is not only extension of the  Islam \& Balbus's (2005) study which is taken into account only parallel viscosity  but also is the extension of the Ferraro's (2007) study which is included parallel viscosity and gyroviscosity only with  $\mathbf{B}=B\mathbf{\hat z}$ magnetic field geometry (also parallel viscosity is negligible in this geometry).
We investigated the nature of gyroviscous instability with the more general magnetic field configuration, i.e., a helical field for  which the parallel viscosity is important. We obtained the dispersion relation of this instability and also derived the  instability criterion. 

The gyroviscous instability shows a complicated dependence on the geometry of the magnetic field through the pitch angle $\theta$. Quataert, Dorland and Hammett (2002) also draw attention to the dependence of the growth rates on the orientation of the magnetic field and the wavevector of the mode. More important than that is the complex coupling between the dynamical effects and the geometry of the magnetic field that makes it almost impossible to single out the net effect of any one agent contributing to stability or instability. But one thing is certain that in its more general case the magnetized, hot and differentially rotating plasmas are always unstable.

Our results show that when only gyroviscosity act on the hot dilute and differentially rotating plasma it brings about instability and growth rates of the instability are higher than MRI (see Fig 1 and 2). In this situation, the finite  $k_{R}$ is destabilized all the wavenumber interval for $30^{\circ}$ in the case of $\mathbf{\Omega}\uparrow\uparrow\mathbf{B}_{z}$ (Fig. 1b and 2a) and for  pitch angles greater than $60^{\circ}$ in the case of $\mathbf{\Omega}\uparrow\downarrow\mathbf{B}_{z}$ (Fig 1d and 2b). Although  parallel viscosity is greater than gyroviscosity, we see that parallel viscosity slightly supresses the unstable modes for the cases of $\mathbf{\Omega}\uparrow\uparrow\mathbf{B}_{z}$ and $\mathbf{\Omega}\uparrow\downarrow\mathbf{B}_{z}$ according to the situation which parallel viscosity is neglected. Also there is an interesting situation: When the angular velocity vector and the $B_{z}$ component of the magnetic field are  parallel ($\mathbf{\Omega}\uparrow\uparrow\mathbf{B}_{z}$), pitch angles smaller than $60^{\circ}$ have maximum growth rates in the great wavenumbers; when the angular velocity vector and the $B_{z}$ component of the magnetic field are  anti parallel ($\mathbf{\Omega}\uparrow\downarrow\mathbf{B}_{z}$), pitch angles greater than $60^{\circ}$ have maximum growth rates in the small wavenumbers (see Fig. 3 and 4). 

In summary, as shown in the figures,  the gyroviscosity effect and pitch angles cause a powerful instability. This gyroviscous instability has a very large gowth rates with $0.5\Omega-3\Omega$ for the different pitch angles. The gyroviscosity instability may be candidate for amplifying very small seed fields. Therefore, this instability may apply to protogalaxies, low-density accretion flows and the intracluster medium of galaxy clusters which the magnetic fields are sufficiently small so that dilute plasma condition is satisfied.

The last but not in the least the dependence of the growth rates on the pitch angle forces  us into a qualititive discussion.   We assumed helical magnetic field geometry. $\mathbf{E}\times\mathbf{B}$ drift generates no current, unless we consider an orbit-averaged $\mathbf{E}\times\mathbf{B}$ wherein the Larmor radii of electrons and ions are different added to this current and viscous flows  generates a net current. Depending on the pitch angles of  current density  and the magnetic field vectors the plasma is either pinched or diluted locally by the Lorentz force, $\mathbf{J}\times\mathbf{B}$. If the pitch angle of the current density vector is smaller than the magnetic field pitch angle the Lorentz force will dilute  otherwise pinch the  plasma locally. It appears that when the relative pitch angles of $\mathbf{J}$ and $\mathbf{B}$ makes the growth rate of the unstable mode higher for all pitch angles range, "the currents caused by FLR effects becomes in phase with the current of the ideal-MHD MRI eigenmode" (Ferraro, 2007).

\section*{Acknowledgments}
We would like to thank the referee for her/his constructive criticism and careful reading the manuscript. We dedicate this work completed in the International Year of Astronomy (IYA) to the honourable memory of Galileo Galilei who had been recognized long since as the founder of the modern science. 
This work is supported by the Ege University Science and Technology Center (EB\.ILTEM) and one of the authors (ED) appreciate the support of the Scientific \& Technological Research Council of Turkey (T\"UB\.ITAK).

\bsp

\label{lastpage}

\end{document}